\documentclass[sigconf]{acmart}

\AtBeginDocument{%
  \providecommand\BibTeX{{%
    \normalfont B\kern-0.5em{\scshape i\kern-0.25em b}\kern-0.8em\TeX}}}

\copyrightyear{2022} 
\acmYear{2022} 
\setcopyright{acmlicensed}\acmConference[CHI '22]{CHI Conference on Human Factors in Computing Systems}{April 29-May 5, 2022}{New Orleans, LA, USA}

\acmBooktitle{CHI Conference on Human Factors in Computing Systems (CHI '22), April 29-May 5, 2022, New Orleans, LA, USA}
\acmPrice{15.00}
\acmDOI{10.1145/3491102.3517706}
\acmISBN{978-1-4503-9157-3/22/04}

\usepackage{xcolor}
\definecolor{myblue}{rgb}{0.3,0.64,0.75}

\usepackage{multirow}
\usepackage{soul}
\newcommand{\new}[1]{{#1}}

\usepackage{amsmath}

\usepackage[utf8]{inputenc}

\begin{document}

\title[Beyond Being Real]{Beyond Being Real: A Sensorimotor Control Perspective on Interactions in Virtual Reality}

\author{Parastoo Abtahi}
\affiliation{
  \institution{Stanford University}
  \streetaddress{353 Serra Mall}
  \city{Stanford}
  \country{USA}}
\email{parastoo@stanford.edu}

\author{Sidney Q. Hough}
\affiliation{
  \institution{Stanford University}
  \streetaddress{353 Serra Mall}
  \city{Stanford}
  \country{USA}}
\email{shough@stanford.edu}

\author{James A. Landay}
\affiliation{
  \institution{Stanford University}
  \streetaddress{353 Serra Mall}
  \city{Stanford}
  \country{USA}}
\email{landay@stanford.edu}

\author{Sean Follmer}
\affiliation{
  \institution{Stanford University}
  \streetaddress{353 Serra Mall}
  \city{Stanford}
  \country{USA}}
\email{sfollmer@stanford.edu}

\renewcommand{\shortauthors}{Abtahi et al.}

\begin{abstract}
    We can create Virtual Reality (VR) interactions that have no equivalent in the real world by remapping spacetime or altering users' body representation, such as stretching the user’s virtual arm for manipulation of distant objects or scaling up the user’s avatar to enable rapid locomotion. Prior research has leveraged such approaches, what we call beyond-real techniques, to make interactions in VR more practical, efficient, ergonomic, and accessible. We present a survey categorizing prior movement-based VR interaction literature as reality-based, illusory, or beyond-real interactions. We survey relevant conferences (CHI, IEEE VR, VRST, UIST, and DIS) while focusing on selection, manipulation, locomotion, and navigation in VR. For beyond-real interaction\new{s}, we describe the transformations that have been used by prior work\new{s} to create novel remappings. We discuss open research questions through the lens of the human sensorimotor control system and highlight challenges that need to be addressed for effective utilization of beyond-real interactions in future VR applications, including plausibility, control, long-term adaptation, and individual differences. 
\end{abstract}



    

\maketitle

\section{Introduction}
    \begin{figure}[t!]
        \centering
        \includegraphics[width=\columnwidth]{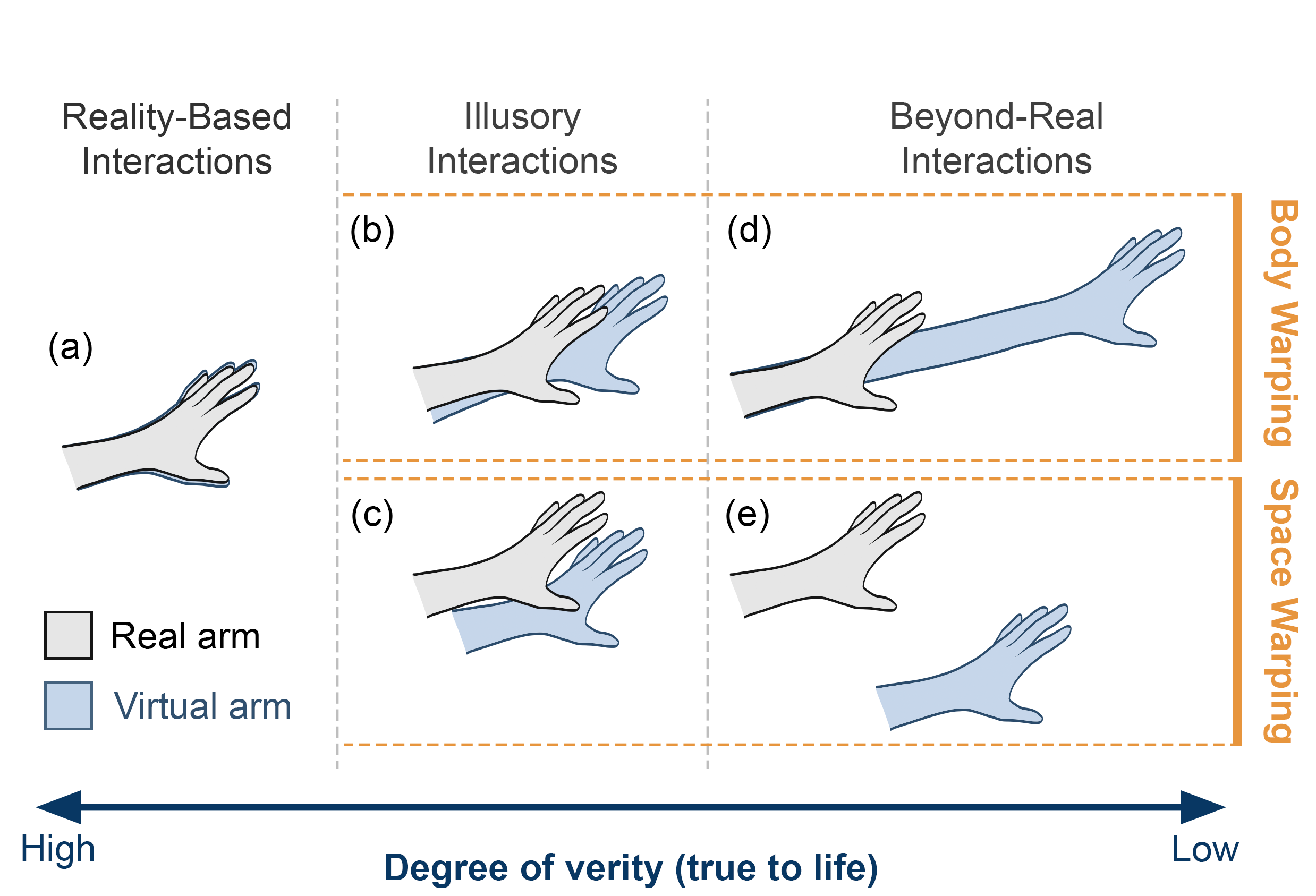}
        \caption{Movement-based VR interactions showcased on a continuum from high to low degree of verity (meaning true to life): reality-based, illusory, and beyond-real interactions, with sensory mismatch created through body-centered (egocentric) or world-centered (allocentric) warping.}
        \label{verity}
    \end{figure}
    
    
    The idea of leveraging VR beyond the replication of reality dates back to the early days of this technology. In a 1965 article, ``The Ultimate Display\new{,}'' Ivan Sutherland proposed that ``there is no reason why the objects displayed by a computer have to follow the ordinary rules of physical reality'' and that ``such a display could literally be the Wonderland into which Alice walked'' \cite{sutherland1965ultimate}. Over the years, other researchers have shared a similar perspective about VR interaction design and have highlighted potential benefits of designing VR interactions beyond reality, including improving human performance \cite{mostafa2014poster} and making interactions more efficient, ergonomic, and accessible \cite{jacob2008reality}. For example, the Go-Go interaction is an arm-extension technique that stretches the user's arm during reach, enabling them to grasp and manipulate distant objects \cite{poupyrev_go-go_1996}. These interactions are designed not to overcome the limitations of VR technology, but to overcome the limitations of our reality.

    In ``Beyond Being There'' (1992)\new{,} Hollan and Stornetta made a parallel argument \new{in response to} telecommunication and computer supported collaborative work \new{research and development at the time}. They argued that when comparing telecommunication to face-to-face communication ``the imitation will never be as good as the real thing. This is true by definition if one is strict in using the old medium as the standard of measurement\ldots requiring one medium to imitate the other inevitably pits strengths of the old medium against weaknesses of the new'' \cite{hollan1992beyond}. They presented a framework around needs, media, and mechanisms, ``to ask the question: what's wrong with (physically proximate) reality?'' and explore new mechanisms that leverage the strengths of the new medium to meet our needs. 
    
    Towards similar goals, we believe there is a need to more systematically investigate what we call beyond-real VR interactions\new{: movement-based} interactions that are not possible in the real world. By \new{``}real world\new{''} we do not imply that there exists an objective physical world independent of the user's subjective mental world. Instead we are referring to what others have called the \new{``}actual world\new{'': what} can be experienced by the human sensory system without the use of technology \cite{lee2004presence}. \new{VR enables full-body interactions with digital content where users can move and act in the virtual world. These interactions do not have to resemble users' movements in the real world, as VR} presents an opportunity to construct \new{imaginative} interactions by remapping users' \new{movements} and altering the resulting sensory feedback. Due to the plasticity of the human sensorimotor system, users have the ability to learn and perform motor tasks under new remappings \cite{ostry2016sensory}. As HCI practitioners, we are interested in exploring VR interactions that are usable and lead to \new{such} motor skill acquisition \new{given} novel dynamics. Thus, we propose describing \new{VR} interactions through the lens of the sensorimotor system, as transformation\new{s} applied to tracking and sensing inputs from the real world. We believe such perspective \new{highlights considerations around action and perception that} are key for understanding the potential, as well as challenges, of beyond-real interactions.
    
    In this paper, we present a framework based on sensorimotor control for categorizing virtual reality interactions as reality-based, illusory, or beyond-real, as shown in Figure~{\ref{verity}}. We further utilize this framing for describing beyond-real interactions as a set of transformations applied to real-world input. We apply the framework to a survey of VR interactions and systematically identify and categorize beyond-real interactions based on their underlying transformations. This survey provides an overview of more than 30 years of beyond-real interaction techniques and identifies key types of transformations that have been explored. Finally, we use the lens of sensorimotor control to map out open research questions central to better understanding the effective use of beyond-real interactions.

    \vspace{0.15cm} 
    \noindent In this work, \textbf{we contribute:}
    \begin{itemize}
        \item \textit{Beyond being real}, a framework based on the human sensorimotor control to describe movement-based VR interactions as transformations applied to input from the real world.
        \item  A literature survey to categorize existing VR interactions (at CHI, IEEE VR, VRST, UIST, and  DIS conferences) as reality-based, illusory, or beyond-real, apply the framework to isolate beyond-real transformations in these selected works, and describe transformation \new{categories} that have been explored by prior research for creating beyond-real interactions.
        \item A discussion of challenges and open research questions that require further investigation \new{of} beyond-real VR interaction design through the lens of sensory integration.
    \end{itemize}
    
\section{Background}
    \new{In VR users can move in virtual spaces and perform full-body interactions.} We focus on \new{these} movement-based VR interactions \cite{gillies2016movement} and action execution in VR (p. 40) \cite{norman2013design}. We approach interaction from a control and optimal behavior perspective \cite{hornbaek2017interaction}, and \new{study} interaction techniques, such as selection, manipulation, and locomotion, that require motor performance \cite{bowman20043d}.
    
    \new{In this section, we first present our categorization of VR interactions as either reality-based, illusory, or beyond-real.} We then provide a high-level background on the human sensorimotor system and control theory. \new{Through this lens, we describe how VR interactions can be thought of as transformations that directly map or remap the user's movements in the real world to renderings in the virtual world.} This insight is central to our work and we believe to research that follows. \new{Here, we use this framing to differentiate between reality-based, illusory, and beyond-real interactions based on whether the transformation applies a remapping and whether that remapping is noticeable by users.} 

    \new{We begin by situating these three categories within the context of prior research.} Thurman and Mattoon \cite{thurman1994virtual} describe different dimensions of VR, including what they call the verity, meaning true to life, dimension. They then use verity to denote ``a continuum of simulation experiences that range from recreations of the physical world as we know it to depictions of abstract ideas which have no physical counterparts\new{.}'' Along this continuum, VR interactions range from interactions with high degree of verity that follow natural laws of the real world to interactions with low degree of verity that follow novel, original laws. Similarly, Slater and Usoh discuss interactions on a spectrum from the mundane to the magical \cite{slater1994body} which map closely to the verity continuum. 
    
    \new{Our three categories} of movement-based VR interactions \new{range from high to low degree of verity}: (1) reality-based interactions that match the user's real-world movements, (2) illusory interactions that create remappings between the user's movements and the virtual renderings that remain unnoticed by users, and (3) beyond-real interactions that create novel remappings between the user's movements and the renderings in the virtual world (see Figure~\ref{verity}).

\subsection{Reality-Based Interactions}
    Highly realistic VR environments that seek to replicate our real-world experiences are used for practical applications, such as training \cite{hill2003virtual}, exposure therapy for treating phobias \cite{rothbaum1997virtual, parsons2008affective}, and post\new{-}traumatic stress disorders \cite{rothbaum2004virtual, hornbaek2017interaction}. These environments also facilitate user interactions that closely resemble interactions in the real world. Jacob et al. proposed the notion of Reality-Based Interactions (RBI) to describe such interactions that employ themes of reality and leverage users' pre-existing knowledge of the everyday, in VR and more broadly \cite{jacob2008reality}. They highlight the benefits of RBI, including accelerated learning, reduced mental effort, facilitated improvisation, and improved performance, particularly in situations involving information overload, time pressure, or stress. They also note that despite the advantages of RBI, designers may explicitly give up realism to gain desired qualities by allowing users to perform many tasks within an application (expressive power) or across different applications (versatility) and to do so rapidly (efficiency), without fatigue or risk of physical injury (ergonomics), and using a varied range of abilities (accessibility). In this work, we focus on VR interactions in which designers explicitly give up realism by creating novel remappings between user inputs and the rendered outputs in VR to overcome the limitations of our experiences in the real world. However, it should be noted that there are many advantages associated with reality-based interactions, and extending interactions beyond reality is not always beneficial, nor is it suitable for all VR applications. 
    
    \subsection{Illusory Interactions}
    As Lanier highlights, our most important canvas in VR is the user's sensorimotor loop \cite{lanier2017dawn}. This technology offers a unique opportunity for manipulating senses, as arbitrary mappings can be created between the user's movements and the rendering of their virtual body. Movement-based VR illusions are remappings that result in a subtle mismatch between the sensory feedback from the virtual system and the sensory feedback from the real world; however, the discrepancy is below the user's perceptual thresholds and is resolved such that the sensory feedback aligns with what the user expects (i.e., the predictions of their internal model). For example, slightly extending the length of the user's arm (Figure~\ref{verity}b) or slightly misplacing the user's hand (Figure~\ref{verity}c) in VR are illusions that will go unnoticed by users. Gonzalez-Franco and Lanier present a model of illusions in VR that describes these processes in more detail \cite{gonzalez2017model}.
    
    Illusions have been explored by researchers to redirect the user's hand while tracing surfaces \cite{kohli2010redirected, zhao2018functional, abtahi2018visuo} or \new{reaching in midair} \cite{azmandian2016haptic, cheng2017sparse, gonzalez2020reach} to provide an improved perceived haptic sensation and overcome the current limitations of VR technology. In these visuo-haptic illusions the mismatch between the visual and proprioceptive feedback is resolved by visual dominance \cite{hecht2009sensory}. Another example of movement-based VR illusions is redirected walking where the rotational movement of the user's head during turns is remapped to a different rotational angle in VR such that their perceived walking path is altered. When studying VR illusions, researchers are concerned with identifying users' perceptual thresholds to ensure that the illusion remains unnoticed \cite{abtahi2018visuo, steinicke2009estimation}. While these illusory interactions are important for improving the perception of realistic (high degree of verity) VR environments, prior research has shown that our cognitive system can adjust to repeated exposure to conflicting stimuli \cite{biocca1998virtual}; thus, there are opportunities for exploration of overt forms of such remapping techniques that go beyond reality. 
    
    \subsection{Beyond-Real Interactions}
    For decades, scholars have emphasized the need for further exploration of virtual experiences beyond replication of reality. In 2003, Schneiderman highlighted that there are many opportunities for enhancing 3D interfaces ``if designers go beyond the goal of mimicking 3D reality'' \cite{shneiderman2003}. In 2005, Casati et al. argued that efforts should be directed towards ``creation of virtual perceptual objects that have no equivalent in the hard reality'' \cite{casati2005subjective}. Gaggioli suggested, in \textit{Human Computer Confluence}, that ``the possible uses of VR range from the simulation of plausible possible worlds and possible selves to the simulation of realities that break the laws of nature and even of logic'' and that VR can provide ``a subjective window of presence into unactualized but possible worlds'' \cite{gaggioli2016human}. Bailenson, in \textit{Experience on Demand}, proposed that the reality bending properties of VR allow us to create experiences ``unbound by the law of the real world, to do impossible things in virtual settings'' and that ``VR is perfect for things you couldn't do in the real world'' \cite{bailenson2018experience}.
    
    From an interaction design perspective, while beyond-real VR interactions can offer benefits, such as making movement-based input more efficient \cite{mostafa2014poster} and ergonomic \cite{jacob2008reality}, they create noticeable incongruencies between the sensory feedback from the real world and the virtual environment. This sensory mismatch has important implications for designing usable beyond-real interactions that people can learn, adapt to, and feel in control of. Therefore, in our work, we carefully consider the human sensorimotor system and approach interaction from a control and optimal behavior perspective \cite{hornbaek2017interaction}. Under this assumption, the human is a goal-directed control system that receives feedback about the state of the world through virtual renderings and behaves so as to change the control signal towards a desired output. The human pursues this goal optimally and adapts to the constraints of the virtual environment. In the next section, we present a simplified model of the human sensorimotor system and optimal control theory that we believe is key in the discussion of beyond-real interactions. We use this theoretical lens throughout the paper to describe beyond-real VR interactions as transformations applied to real-world input. \new{We conduct a survey of beyond-real transformations that have been utilized by prior research} and highlight open research questions that remain in the design and evaluation of usable beyond-real VR interactions. 
   
    \subsection{Sensorimotor System and Control Theory}
    Human performance may be modelled at various levels of behavior: skill-based, rule-based, and knowledge-based behaviors \cite{rasmussen1983skills}. Optimal Feedback Control (OFC) theory focuses on skill-based behavior (e.g., catching a ball) and has been used to predict how the human brain plans and controls movement \cite{scott2004optimal} by studying the link between high-level goals and real-time sensorimotor control strategies \cite{todorov2004optimality}. This theory suggests that the Central Nervous System (CNS) acts as a feedback controller, continuously converting sensory input into motor output \cite{ueyama2014mini} and it does so optimally, based on a performance metric, such as obtaining minimal uncertainty in the state estimate \cite{van2002role}. Researchers have also proposed using a Mini-Max Feedback Control (MMFC) model, an extension of the OFC model that minimizes energy consumption under the assumption of worst-case uncertainty \cite{ueyama2014mini}.
    
    
    \begin{figure*}[t!]
        \centering
        \includegraphics[width=0.85\linewidth]{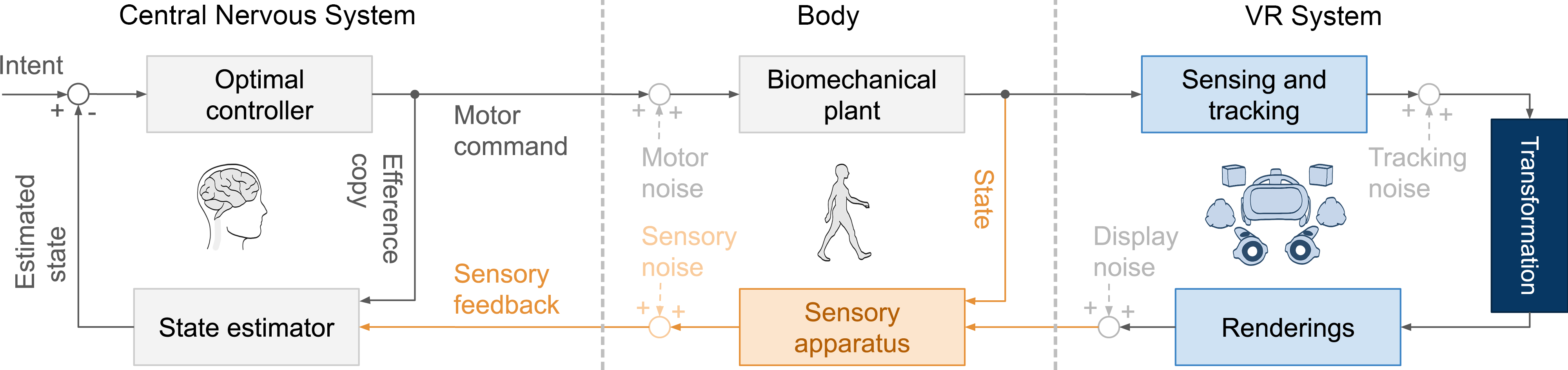}
        \caption{Flow of control signals in movement-based interactions through the central nervous system, body, and VR system.}
        \label{control_model}
    \end{figure*}
    
    
    \subsubsection{Overview}
    Figure~\ref{control_model} shows how the CNS interacts with the body and the VR system during movement-based interactions. In this diagram blocks represent key components, and arrows denote the flow of control signals, clockwise from the top left. The optimal controller outputs motor commands based on the discrepancy between the desired and estimated states \cite{wolpert2000computational}. These motor commands lead to movements in the real world that are then subject to body dynamics and the effects of the environment, such as external forces. The VR system includes sensing and tracking devices that capture the user's movements. Movement-based VR interactions can be thought of as transformations applied to these signals captured from the real world. In reality-based interactions the transformations create a 1:1 mapping to the VR renderings\new{;} in illusory interactions the transformations create subtle remappings that are unnoticed by users\new{;} and in beyond-real interactions the transformations create novel remappings. The human sensory apparatus receives sensory feedback from both the real world and the virtual system (shown in orange). The state estimator receives the sensory feedback through the sensory apparatus as well as an efference copy of the original motor signal \cite{bhushan1999evidence}. While noise is present at all stages \cite{van2002role}, we only show discrete examples in this diagram.
        
    \begin{figure}[b!]
        \centering
        \includegraphics[width=0.9\columnwidth]{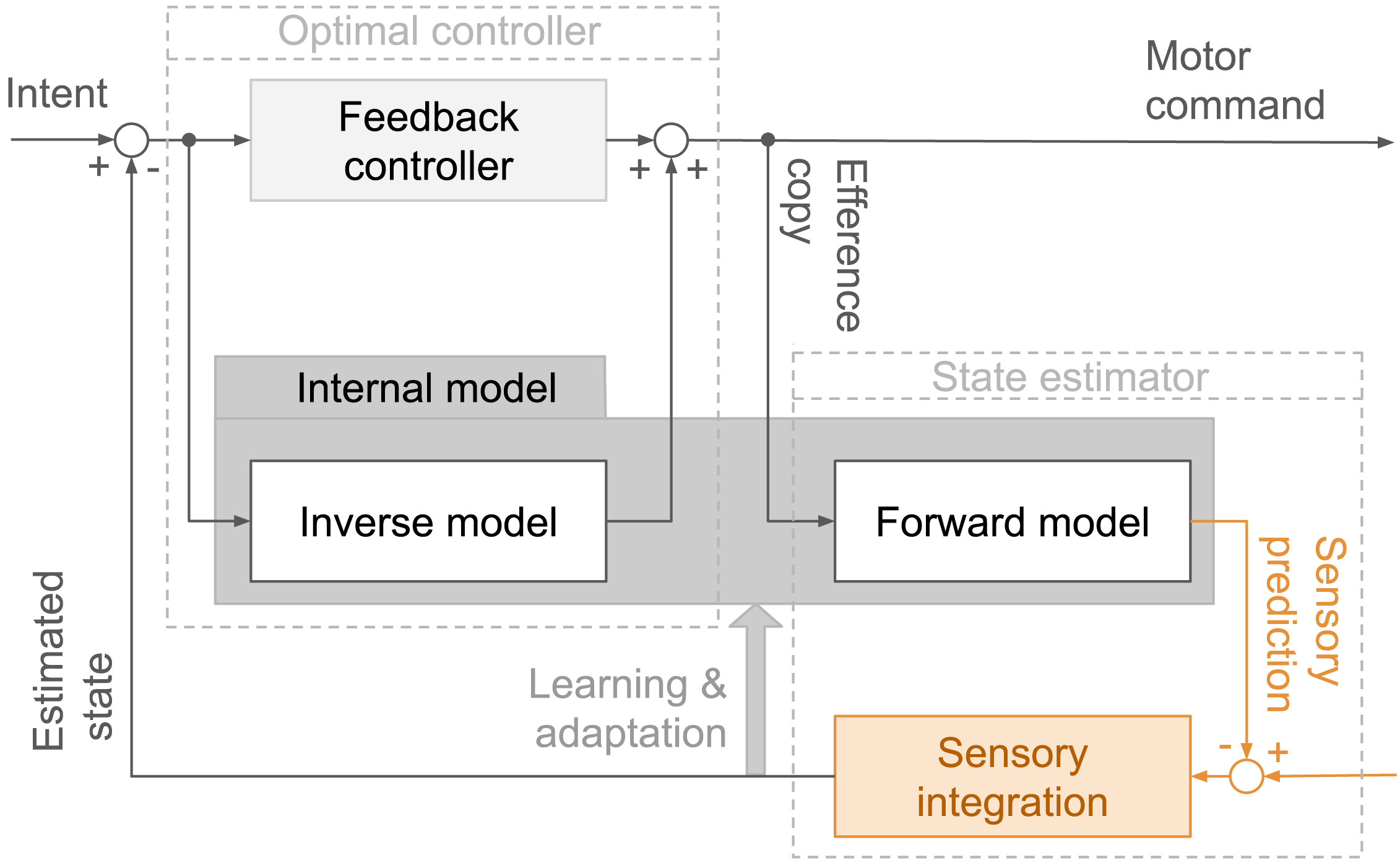}
        \caption{Control signals in the central nervous system.}
        \label{brain}
        \vspace{-0.25cm}
    \end{figure}
    

    \subsubsection{Central nervous system} Figure~\ref{brain} shows the subcomponents of the CNS and the flow of control signals. The feedback controller outputs motor commands based on the discrepancy between the desired and estimated states, which is then combined with the output of an adaptive inverse model \cite{wolpert2000computational}. An efferent copy of motor signals is sent to a forward model that predicts the result of motor commands \cite{bhushan1999evidence}. The forward and inverse models are collectively referred to as the internal model and capture information about the context and the properties of the sensorimotor system.  
    
    \subsubsection{Sensory integration}
    Multisensory integration is a complex process that modifies the original signal based on low-level sensory information, top-down influences of the internal model, and a range of cognitive factors; therefore, it is perhaps more accurately described as multisensory interaction \cite{talsma2015predictive}. This interaction is task-dependent and may be affected by the modality of the stimulus as well as the information content of the feedback \cite{sober2005flexible}. Multisensory processing is also influenced by attention \cite{talsma2015predictive} and human emotional responses to stimuli \cite{schreuder2016emotional}. Finally, the central nervous system minimizes uncertainty by refining sensory signals based on prior knowledge and memory \cite{ueyama2014mini}. 
    
    
    Redundancy in the sensorimotor system ensures robustness \cite{flanagan2003prediction} such that elimination of a feedback has minor effects on motor behavior \cite{doya2008modulators}; however, perturbations of the same signal may significantly alter movement \cite{cruse1990utilization}. For example, while reaching without sight results in minor errors, visual distortions have been shown to lead to drastic compensatory movements \cite{saunders2005humans, schoner1994dynamic}. Therefore, sensory integration predominantly addresses unexpected changes based on prediction errors \cite{talsma2015predictive}. Note that OFC theory is concerned with errors that are referred to as \emph{slips}, and not \emph{mistakes} that arise from incorrect intentions (p. 414) \cite{norman1986user}. 

    \subsubsection{Learning and adaptation}
    Prediction errors drive simultaneous perceptual and motor learning \cite{ostry2016sensory, flanagan2003prediction}. While both the forward and inverse models are adapted \cite{wolpert2000computational}, it has been shown that prediction learning precedes the learning of new control policies \cite{flanagan2003prediction}. Beyond adaptation to perturbations, humans can learn to synthesize movement under entirely novel dynamics \cite{haith2013model}. An example of sensorimotor learning is prism adaptation in which an individual performs perceptual motor tasks while wearing goggles that shift their visual field \cite{rossetti1998prism}. An interesting characteristic is that the effects of adaptation after removing the goggles, known as aftereffects, are not global, and only result in a systematic movement bias for the specific, practiced task \cite{arbib2009tool}.
    
    \vspace{0.15cm} 
    \noindent We use this theoretical background to first present a descriptive framework for beyond-real interactions and then discuss open research questions and challenges that remain in the context of sensory integration and adaptation. 
    
    \begin{figure*}[t!]
        \centering
        \includegraphics[width=\linewidth]{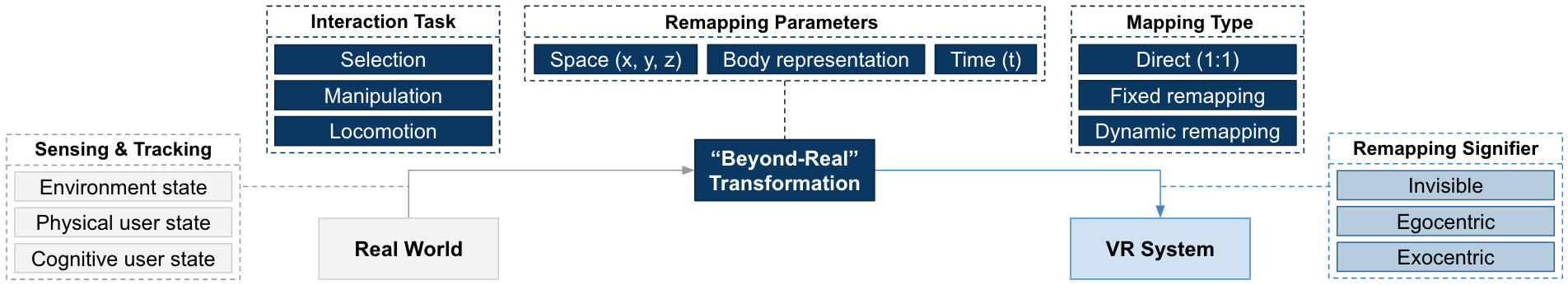}
        \caption{The VR system receives input from the real world, applies beyond-real transformations, and renders the remapping.}
        \label{framework}
        \vspace{-0.15cm}
    \end{figure*}

\section{The Beyond Being Real Framework} \label{sec:framework}
    We present \textit{beyond being real}, a framework \new{using a sensorimotor control perspective} for investigating movement-based VR interactions that have no equivalent in the real world. The framework, shown in Figure~\ref{framework}, provides a scaffolding for describing beyond-real interactions in three stages, by describing the sensing and tracking data in the real world, the set of transformations applied to the real-world input for creating novel remappings in VR, and the VR renderings that provide signifiers and feedback to improve the usability of beyond-real interactions. In this section, to highlight this descriptive power, we return to the example of the Go-Go interaction technique \cite{poupyrev_go-go_1996}. Note that while we provide a high-level description of the Go-Go interaction here, the framework is better \new{suited to} describing a specific implementation of the interaction technique that includes more details. \new{For a more in-depth walk-through example, please refer to Appendix B.}
    
    
    \subsection{Real-World Sensing and Tracking}
    The virtual system has limited information about the state of the real world. For example, in most commercial VR systems the position of the headset is known; however, the system often does not have direct knowledge of the user's body pose. The real-world information often describes the environment state (e.g., room dimensions, obstacles), users' physical state (e.g., head position, hand pose), and their cognitive state (e.g., attention, workload). In the first stage of the framework, we identify what sensing and tracking devices are available, what limitations they have (e.g., range, rate, accuracy), and what real-world data from those devices is being utilized by the virtual system. In the Go-Go interaction technique, the inputs from the real world are the user's physical arm length ($d$) and the user's real hand position ($\vec{H_{r}}$) in the user's egocentric frame of reference. In this case, the origin is the user's chest position, approximately determined based on the position of the VR headset.
    
    \subsection{Beyond-Real Transformations} 
    The input passed on to the virtual system is then either disregarded, mapped directly, or remapped. Remappings can be fixed throughout the interaction or may dynamically change based on users' actions captured by the input data \cite{latta1994conceptual}. These mappings can be described as transformation\new{s} that take input from the real world and modify \new{the space, the user's body representation, or time parameters}. 
    
    In the second stage of the framework, \new{we focus on the goal of the beyond-real interaction (i.e., interaction task). To describe the} sets of transformations applied to the real-world data, we identify what parameters are modified as a result of the remappings (space, body, or time) and what the mapping type is (direct, fixed, or dynamic). In the Go-Go interaction, the beyond-real transformation is a dynamic remapping that modifies the user's body representation \new{for manipulation of distant objects}. More specifically, the user's virtual hand ($\vec{H_{v}}$) is extended during the last 1/3 of their reach range, for a given coefficient k ($ 0 < k < 1 $): 
	\begin{align*}	
        \vec{H_{v}} = \left\{	
        \begin{array}{ll}	
        \vec{H_{r}} & \mbox{if  $\|\vec{H_{r}}\| < \frac{2}{3}d$}\\	
        \vec{H_{r}} + k (\|\vec{H_{r}}\| - \frac{2}{3}d)^2 & \mbox{otherwise}	
        \end{array}\right.	
    \end{align*}
    
    \noindent While our framework focuses on VR interactions, such transformations are not unique to VR and may be used to describe remappings between movement-based user inputs and outputs of computing systems more broadly. For example, by modifying the control/display ratio, the movements of the mouse can be remapped to the movements of the cursor. \new{However, for beyond-real VR interactions specifying these} transformations can be \new{especially useful for determining} the sensory mismatches that users experience as a result of the remapping, which we discuss in more detail in section~\ref{sec:research_questions}.
    
    \subsection{VR Renderings and Remapping Signifiers}
    The virtual system renders information through output devices, such as the VR headset and headphones. While some renderings are mapped to the input (i.e., they directly result from user actions), others are independent of the real-world input and the applied transformations. Independent renderings may communicate to users what mappings exist, prior to the execution of actions. These signifiers may be egocentric (e.g., visible features of the body representation) or exocentric (e.g., features in the environment or specialized objects). The concept of ``User Representation'' defined by Seinfeld et al. \cite{seinfeld2021user} is closely related to egocentric signifiers. More specifically, User Representations are virtual elements that extend users' physical bodies and they ``may have signifiers that communicate the actions they support.''
    
    In the third stage of the framework, we identify the aspects of the renderings that are independent of the real world and communicate the remapping to users (invisible, egocentric, \new{or} exocentric signifiers). Remapping signifiers have important implications for learnability and adaptation to novel remappings (see discussion in section~\ref{sec:research_questions}). In the Go-Go interaction, there are no visible signifiers that communicate the remapping to users independent of the user's actions. Therefore, users can only discover the remapping after they extend their arm more than 2/3 of their arm length.
    
    \new{In the following section, we present a survey of beyond-real interactions previously presented at HCI conferences. We apply this framework to those interactions to isolate and group the beyond-real transformations that have been explored by prior works.}

\section{Survey of Beyond-Real Interactions}
    We conducted a systematic review of literature, following PRISMA guidelines \cite{moher2009preferred}, to (1) understand past research trends with respect to reality-based, illusory, and beyond-real movement-based VR interactions, (2) evaluate to what extent VR interaction research has explored beyond-real transformations, and (3) \new{explore} whether or not researchers have considered the human sensorimotor loop in their exploration of beyond-real interactions.  
    
    In this survey, we focused on action execution and more specifically, on motor performance (p. 40) \cite{norman2013design}. 3D interaction techniques have been categorized into selection, manipulation, wayfinding, locomotion, system control, and symbolic input \cite{bowman20043d}. We were particularly interested in \emph{selection}, \emph{manipulation}, and \emph{locomotion}, as they require users to act on the world. We chose to exclude \emph{symbolic input} and \emph{system control} through which users change the mode or state of the system, as they do not have a counterpart in our physical reality and fall outside the scope of our work.
    
    Navigation is conceptualized as having two components: \emph{wayfinding} refers to the cognitive component of navigation and \emph{locomotion} describes the movement from one place to another \cite{montello2006human}. Due to its cognitive nature, wayfinding cannot be \new{fully} captured through the lens of sensorimotor control. While we use \new{the term} ``navigation'' as part of our search query to capture all locomotion papers, we do not focus on works that only address wayfinding. 
    
    
\subsection{Method}
    
    \subsubsection{Phase 1: Identification}
    We searched the ACM Digital Library for full papers targeting the following venues: the ACM Conference on Human Factors in Computing Systems (CHI), the ACM Symposium on Virtual Reality Software and Technology (VRST), the ACM Symposium on User Interface Software and Technology (UIST), and the ACM Conference on Designing Interactive Systems (DIS). Additionally, we searched IEEE Xplore targeting the IEEE Conference on Virtual Reality and 3D User Interfaces (IEEE VR). We focused our search on VR interaction techniques, allowing terms for common interaction techniques focused on action (selection, manipulation, locomotion, navigation) to appear in either the title or abstract. As we sought to understand trends of research attention to reality-based, illusory, and beyond-real interactions over time, papers placed in our identification phase date back to 1988. \new{Note that we did not use keywords in our search query}. An example of the way our queries were structured: \vspace{0.1cm}
\begin{verbatim}
Title:((interaction* OR select* OR manipulat* OR
locomot* OR navigat*) AND (virtual OR VR)) OR
Abstract:(((interaction* OR select* OR manipulat* OR 
locomot* OR navigat*) AND (virtual OR VR))
\end{verbatim}
    \noindent
    This phase found a total of \new{1268} full papers for further screening.

    \subsubsection{Phase 2: Screening}
    We excluded papers that were not focused on virtual reality (\new{326}). This excluded augmented reality and other non-immersive platforms such as tabletop displays. Furthermore, we excluded papers that were not focused on interaction techniques (\new{271}). We defined interaction techniques as means by which the user engages with the virtual content \new{through movement} - as opposed to novel infrastructure, rendering techniques, collision detection algorithms, \new{visualizations,} descriptions of input \new{devices} or haptic devices. Screening reduced our set to \new{671} papers.
    
    \begin{figure}[h!]
        \centering
        \includegraphics[width=0.85\columnwidth]{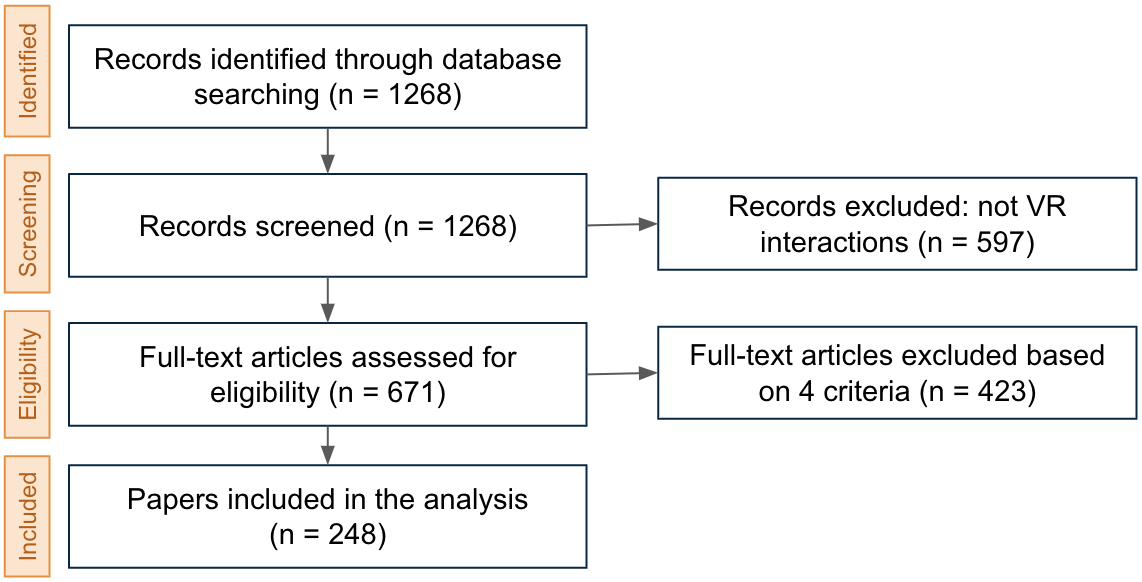}
        \caption{Flow of information through the different phases of our systematic review, following PRISMA guidelines.}
        \label{PRISMA}
        \vspace{-0.15cm}
    \end{figure}
    
    \subsubsection{Phase 3: Eligibility}
    We excluded papers that were analyses or experimental evaluations of existing interaction techniques (283), applications of interaction techniques to real-world problems (127), surveys of interaction techniques (8), and revisions of the same interaction techniques produced by the same authors (5). This step built our final study set of \new{248} papers.
    
    \subsubsection{Phase 4: Dataset and coding}
    We coded each included paper along four dimensions: (1) type of interaction: reality-based, illusory, or beyond-real, \new{and if beyond-real} (2) interaction task: selection, manipulation, locomotion, or wayfinding, (3) \new{remapping parameter}: space, time, or body, and (4) \new{consideration} of sensorimotor loop: yes/no. Often papers developed techniques that leveraged multiple transformations and could be applied to multiple interaction tasks. We applied multiple labels in these cases. Note that while we have coded for all interaction tasks that appeared in our dataset, we \new{focus on interactions that require skilled motor actions and not higher-level cognition} (e.g., wayfinding) in the results. 
    

\subsection{High-Level Survey Findings}

\subsubsection{Interaction types}

\begin{figure}[h!]
        \centering
        \includegraphics[width=1\columnwidth]{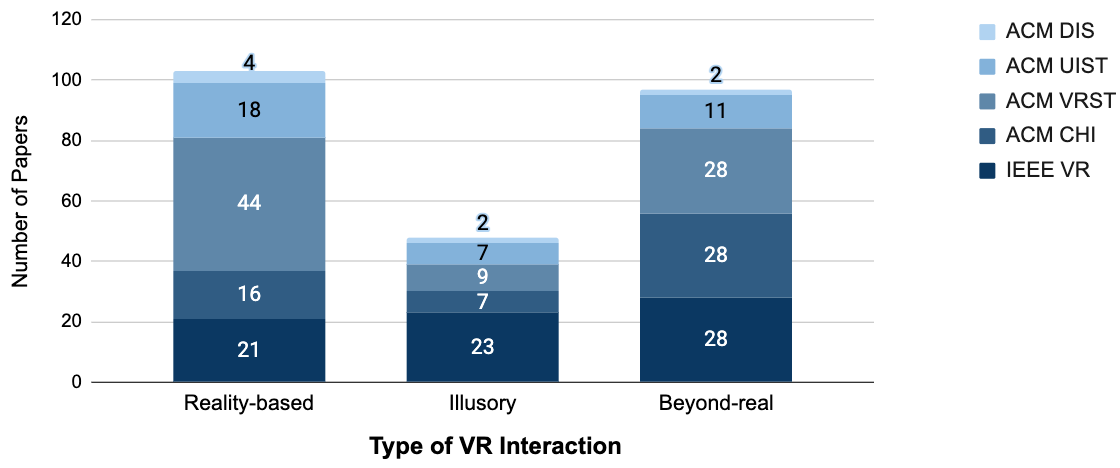}
        \caption{Bar chart partitioning VR interaction papers into reality-based, illusory, and beyond-real categories.}
        \label{interaction_types}
        \vspace{-0.15cm}
    \end{figure}

Of the interaction techniques we analyzed, we found: \new{103} reality-based (\new{42}\%), \new{48} illusory (\new{19}\%), and 97 beyond-real (\new{39}\%), as shown in Figure~\ref{interaction_types}. While the frequencies of beyond-real and reality-based interaction papers remained relatively consistent over time, we saw a jump in the \new{number} of illusory interactions \new{after} 2016; \new{12 illusory interactions were presented before (in 20 years, 1996-2016) and 36 after (in 4 years, 2017-2021). For a full list of illusory interaction papers, please refer to Appendix A.}



    
\subsubsection{Interaction tasks}

\begin{figure}[h!]
        \centering
        \includegraphics[width=1\columnwidth]{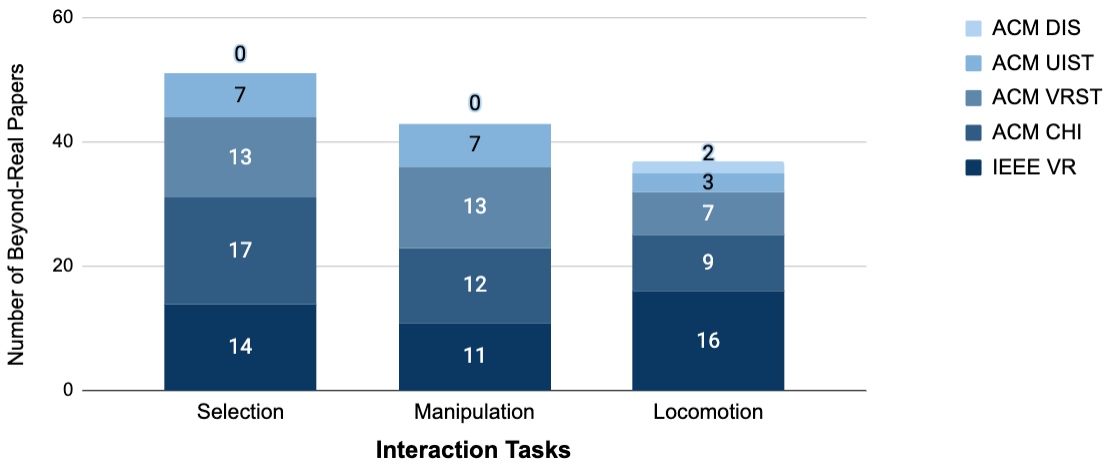}
        \caption{Bar chart partitioning beyond-real interaction papers based on the interaction task they focus on.}
        \label{interaction_tasks}
        \vspace{-0.35cm}
    \end{figure}

Of the \new{97} beyond-real interaction tasks, we found: \new{51} selection (\new{39}\%), \new{43} manipulation (\new{33}\%) and \new{37} locomotion (\new{28}\%), shown in Figure \ref{interaction_tasks}. These numbers do not sum to \new{97} because, as mentioned, some interaction techniques are multi-purpose. For example, beyond-real techniques that leverage miniature reconstructions of virtual environments allow for both selection and manipulation of occluded objects \cite{li_vmirror_2021}. 

\subsection{Beyond-Real Transformations Explored}


\begin{figure}[b!]
    \centering
    \includegraphics[width=\columnwidth]{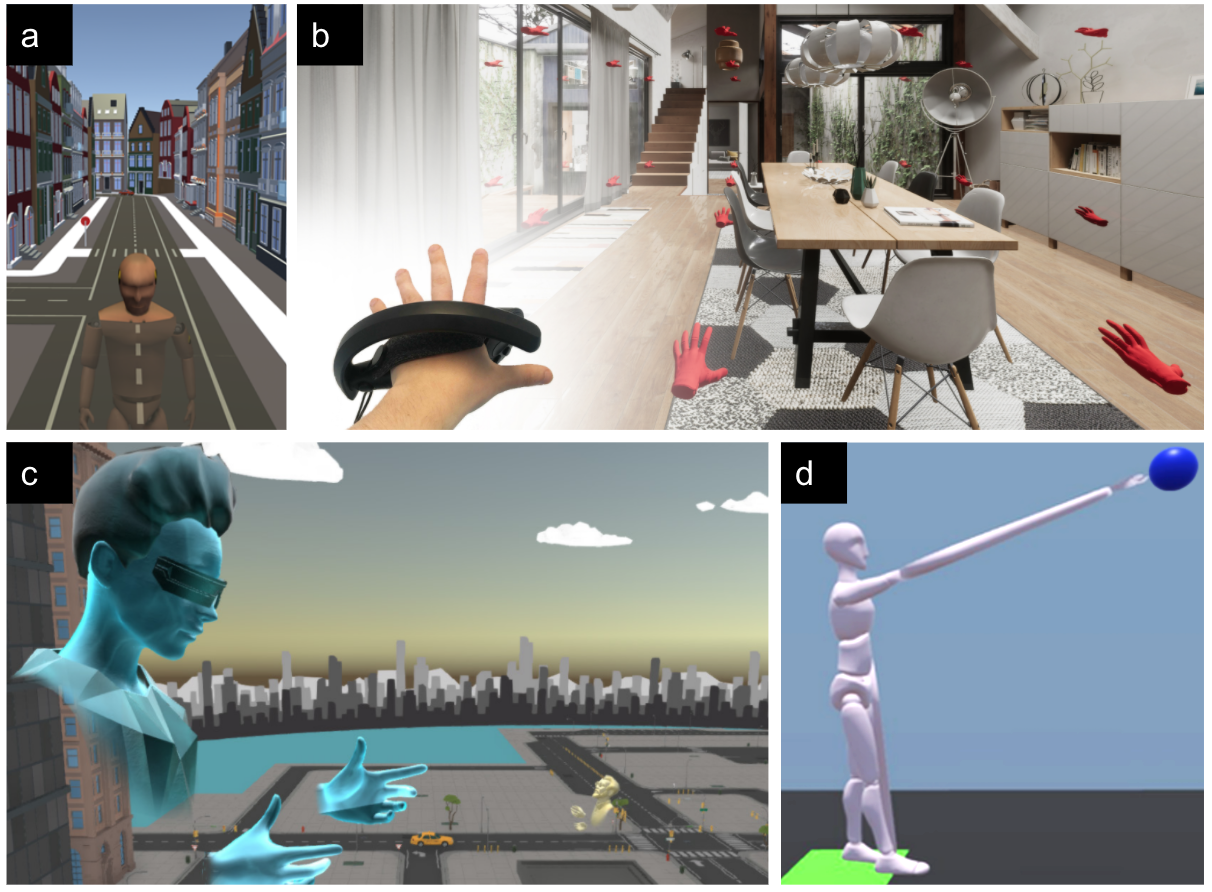}
    \caption{Examples of beyond-real interaction techniques in the literature survey (a: \cite{abtahi_beyond_2019}, b: \cite{schjerlund_ninja_2021}, c: \cite{xia_spacetime_2018}, d: \cite{mcintosh_iteratively_2020}).}
    \label{collage}
\end{figure}

Here we focused on the subset of papers in our survey that explore beyond-real interactions. We applied our framework (described in section~\ref{sec:framework}) to these prior works in order to isolate the beyond-real transformations they utilize and organized these transformations into three groups (space, body, or time) \new{based on their remapping} parameter. Using a combination of inductive and deductive coding, we then identified subcategories of transformation\new{s} that we describe in this section. It should be noted that these transformations can be described in multiple ways. Additionally, a user may reason about these transformations differently than how the transformations are implemented \new{in practice}. Therefore, the choice of transformation functions may depend on the specifics of the design, context of the interaction, or the aim of the analysis.

\subsubsection{Space Transformations} Space transformations create remappings \new{of movement} in 3D space.

\vspace{0.15cm}
\noindent Translation: Prior work has explored space translation for locomotion tasks in VR, specifically to augment walking. Translational gain amplifies the shift of the virtual ground under the user's feet \new{to enable users to walk more rapidly in VR} \cite{wilson_object_2018, rietzler_telewalk_2020}. 



\vspace{0.15cm}
\noindent Scaling: The scale of rendered content can be altered in VR to enable novel forms of interactions with virtual objects and the surrounding environment. For example, users can scale down the environment to obtain a high-level overview \new{and more easily manipulate large virtual objects} \cite{xia_spacetime_2018}. Scale-based remappings have also been leveraged for locomotion. For example, the virtual world can be scaled down with the center of scaling at the midpoint between the user’s eyes, allowing them to walk rapidly through a miniature world \cite{abtahi2019m}.

\vspace{0.15cm}
\noindent Duplication: A fixed mapping \new{can be} created between the real-world space and multiple copies of the environment in VR. Poros is one such interaction technique that displays proxies of remote regions of the virtual environment \cite{pohl_poros_2021}. The changes made to the miniature VR space propagate to the full-size virtual environment, enabling users to utilize the proxy to manipulate objects. Another technique, vMirror, uses strategically placed mirrors and their reflections to allow users to manipulate obscured objects \cite{li_vmirror_2021}. 



\begin{table*}[t!]
    \caption{Beyond-Real Interactions Categorized by Transformation Type} \label{tab:transforms} 
    \begin{tabular}{l l l}
         \toprule
         \textbf{Transformation} & \textbf{Count} & \textbf{References} \\
        \midrule
         \textbf{Space} & 44 & \\
        \midrule
        Translation & 18 & \cite{wilson_object_2018, abtahi_im_2019, wolf_jumpvr_2020, von_willich_podoportation_2020, rietzler_telewalk_2020, wartell_third-person_1999, song_nonlinear_1993, mirhosseini_automatic_2017, jacob_habgood_rapid_2018, weissker_multi-ray_2019, eroglu_design_2021, englmeier_spherical_2021, elmqvist_balloonprobe_2005, veit_dynamic_2010,louvet_combining_2016, mossel_large_2016, nuernberger_evaluating_2017, lee_evaluating_2020} \\
                    Scaling & 27 & \cite{abtahi_im_2019, von_willich_podoportation_2020, pohl_poros_2021, li_vmirror_2021, wartell_third-person_1999, song_nonlinear_1993, pierce_navigation_2004, kopper_design_2006, cashion_dense_2012, suma_impossible_2012, montano-murillo_slicing-volume_2020, eroglu_design_2021, englmeier_spherical_2021, elmqvist_balloonprobe_2005, veit_dynamic_2010, veit_cros_2012, debarba_embodied_2015, louvet_combining_2016, argelaguet_giant_2016, lee_evaluating_2020, cirio_magic_2009, montano_murillo_navifields_2017, montano-murillo_drift-correction_2019, sra_your_2018, wolf_jumpvr_2020, xia_spacetime_2018, yu_magnoramas_2021} \\
        Duplication & 18 & \cite{pohl_poros_2021, li_vmirror_2021, pierce_navigation_2004, freitag_interactive_2018, nam_worlds--wedges_2019, mirhosseini_exploration_2019, montano-murillo_slicing-volume_2020, yu_magnoramas_2021, xia_spacetime_2018, wang_slice_2020, stoev_application_2002, chittaro_navigation_2006, monclus_virtual_2009, argelaguet_visual_2009, zhang_virtual_2020, schott_virtual_2020, tseng_facewidgets_2019, xia_spacetime_2018}\\
                 \midrule
        \textbf{Body} & 55 & \\
        \midrule
        Alternative Morphologies & 8 & \cite{abtahi_im_2019, schjerlund_ninja_2021, fairchild_heaven_1993, wang_object_2015, poupyrev_go-go_1996, mcintosh_iteratively_2020, griffin_out--body_2019, xia_spacetime_2018} \\
        Movement Remapping & 29 & \cite{fukatsu_intuitive_1998, balakrishnan_exploring_1999, tanriverdi_interacting_2000, song_developing_2000, fairchild_heaven_1993, zeleznik_pop_2002, song_nonlinear_1993, frees_precise_2005, dominjon_haptic_2006, terziman_shake-your-head_2010, song_handle_2012, slambekova_gaze_2012, tregillus_vr-step_2016, caputo_smart_2017, mirhosseini_automatic_2017, sra_breathvr_2018, teng_aarnio_2019, xu_clench_2019, sidenmark_eyeamphead_2019, hanson_improving_2019, hayatpur_plane_2019, sidenmark_outline_2020, freiwald_walking_2020, yu_gaze-supported_2021, sidenmark_radi-eye_2021, ahn_stickypie_2021} \\
                & & \cite{wendt_gud_2010, xia_spacetime_2018, zhang_perch_2019}\\
        Tool Use & 22 & \cite{schkolne_surface_2001, park_select_2012, tu_crossing-based_2019, prouzeau_scaptics_2019, jiang_handpainter_2021, zeleznik_pop_2002, stuerzlinger_efficient_2002, jackson_yea_2018, mardanbegi_eyeseethrough_2019, wang_scene-context-aware_2021, sturman_hands-interaction_1989, arora_magicalhands_2019, song_developing_2000, de_haan_towards_2002, argelaguet_overcoming_2008, meyrueis_d3_2009, stenholt_efficient_2012, wonner_starfish_2012, louvet_combining_2016, lu_investigating_2020, baloup_raycursor_2019, funk_assessing_2019}\\
        \midrule
        \textbf{Time} & 4 & \\
        \midrule
         Time Travel & 2 & \cite{simon_active_2007, xia_spacetime_2018} \\
         Speed Change & 2 & \cite{rietzler_matrix_2017, jiang_mediated-timescale_2020} \\
        \bottomrule
    \end{tabular}
\end{table*}

\vspace{-0.2cm}
\subsubsection{Body Transformations} Beyond-real interactions may alter users' body representation, which has been defined in a variety of ways. Given our focus on action, in this paper, we refer to the categorization of bodily representation by Martel et al. \cite{martel2016tool} which includes \emph{body image}, \emph{body structural description}, and \emph{body schema}. \emph{Body image} relies heavily on visual input and refers to the conscious representation of the body's shape and size. \emph{Body structural description} is a conscious spatial map of the body parts and their relationships, informed primarily by somatosensory and visual systems. \emph{Body schema} refers to the unconscious and highly plastic representation of the body parts, including posture, shape, and size. It should be noted that while some beyond-real interactions may be described in other ways, such as space transformations, they are perceived by users as transformations of body representation. Due to the plasticity of body representation in the human brain, this egocentric perspective is necessary to more accurately capture users' expectations and actions.

\vspace{0.15cm}
\noindent Alternate Morphologies: In VR, users can embody avatars with novel \new{sizes,} forms, and structures. For example, Ninja Hands maps the movement of a single hand to multiple hands to ease distant target selection \cite{schjerlund_ninja_2021}. Another paper iteratively adjusts the length, and therefore range of motion, of the avatar's forearms and fingers to achieve better performance on specific tasks \cite{mcintosh_iteratively_2020}. \new{Note that users may perceive space scaling transformations as body scaling and a form of alternate morphology.}

\vspace{0.15cm}
\noindent Movement Remapping: Movement of the user's body in the real world can be altered virtually to represent another type of movement. 
Shake-Your-Head maps lateral and vertical head movements to walking and jumping, thus enabling in-place locomotion \cite{terziman_shake-your-head_2010}. Walking by Cycling maps real-world pedaling motions to the walking of a virtual avatar \cite{freiwald_walking_2020}. \new{Movement remappings} can also be used for object selection and manipulation. \new{For example,} users can move and rotate objects by grasping an imaginary handle bar skewering virtual objects \cite{song_handle_2012}. One prominent form of movement remapping comes in the form of gaze-based interactions, which translates eye movement to hand input such as grabbing or shifting objects \cite{yu_gaze-supported_2021}.


\vspace{0.15cm}
\noindent Tool Use: Increasingly, evidence is emerging that tool-use may affect body image \cite{martel2016tool} and body schema \cite{cardinali2009tool}. This point is further discussed by Seinfeld et al. \cite{seinfeld2021user} through the concept of User Representations. Therefore, in some cases it may be appropriate to describe tool-based interactions as a body representation transformation. For example, ray-casting is a popular tool-based selection technique in which a light ray extends from the user's finger and intersects with various objects. \new{Ray-casting can be enhanced to select the nearest target} \cite{baloup_raycursor_2019}. For multiple object selection, researchers have also developed techniques that map the position of users' hands to virtual brushes and lassos \cite{stenholt_efficient_2012}.


\subsubsection{Time Transformations} \new{Remappings can be created} that alter the user's perception of and interaction\new{s} with time.

\vspace{0.15cm}
\noindent Time Travel: In \new{VR} it is possible to allow the user to visit the future or retrace their temporal footsteps. One technique allows the user to revisit old checkpoints along a path \new{for navigation} \cite{retracepaths}. The users' timeline of engagements with the \new{VR} application is recorded and becomes another dimension along which they may travel.


\vspace{0.15cm}
\noindent Speed Change: Users of virtual reality applications can develop skills with gentler learning curves with the help of time manipulation - for instance, slower motion of a tennis ball so that beginner players can successfully return it \cite{tennisball}. The motion of avatars in VR can also be accelerated or decelerated to change the user's perception of time \cite{matrixhasyou}.



\subsection{Survey Results for Beyond-Real Interactions}

\subsubsection{Transformation types} 
    \noindent Of the \new{97} beyond-real transformation papers, we found: \new{44} space transformations (\new{45}\%), \new{55} body transformations (\new{57}\%), and \new{4} time transformations (\new{4}\%). All of the beyond-real interactions surveyed are shown in Table \ref{tab:transforms}, where they are \new{organized based on subcategories of transformations}. Of space transformations, we found \new{27} are scaling (\new{61}\%), \new{18} are translation (\new{41}\%), and \new{18} are duplication (\new{41}\%). Of body transformations, we found \new{22} involve tool use (\new{40}\%), \new{29} are movement remapping (\new{53}\%), and \new{8} are alternative morphologies (\new{14}\%). Of time transformations, 2 leveraged speed change and 2 used time travel. Multiple space and body transformations consisted of sub-transformations. For example, techniques that scale the user's jumps scale the environment (shrink it to make the jump appear higher) as well as translate it (have the ground move faster while the user is in the air).
    
\subsubsection{Consideration of sensorimotor loop} We found that \new{23} of the \new{97} beyond-real papers consider the effect of sensory conflict (\new{24}\%); only \new{4} of these were published before \new{2017}. Usually discussions of the sensorimotor loop center around simulator sickness evaluated on study participants with the standard Simulator Sickness Questionnaire (SSQ). Primarily SSQ scores are one of several metrics, such as frustration or movement instability, used to assess the effectiveness of a given interaction technique. We found limited examination of causal factors in favor of a more empirical treatment. Deeper model-based analysis such as that enabled by control theory may position researchers to design interaction techniques that do not induce simulator sickness at the outset, as well as iterate more efficiently upon recognition of factors responsible for sensory mismatch. The results of this survey additionally suggest a strong opportunity for the \new{VR} community to explore sensorimotor issues with interaction techniques beyond simulator sickness.

\section{Open Research Questions}
\label{sec:research_questions}

    \begin{figure}[h]
        \centering
        \includegraphics[width=1\columnwidth]{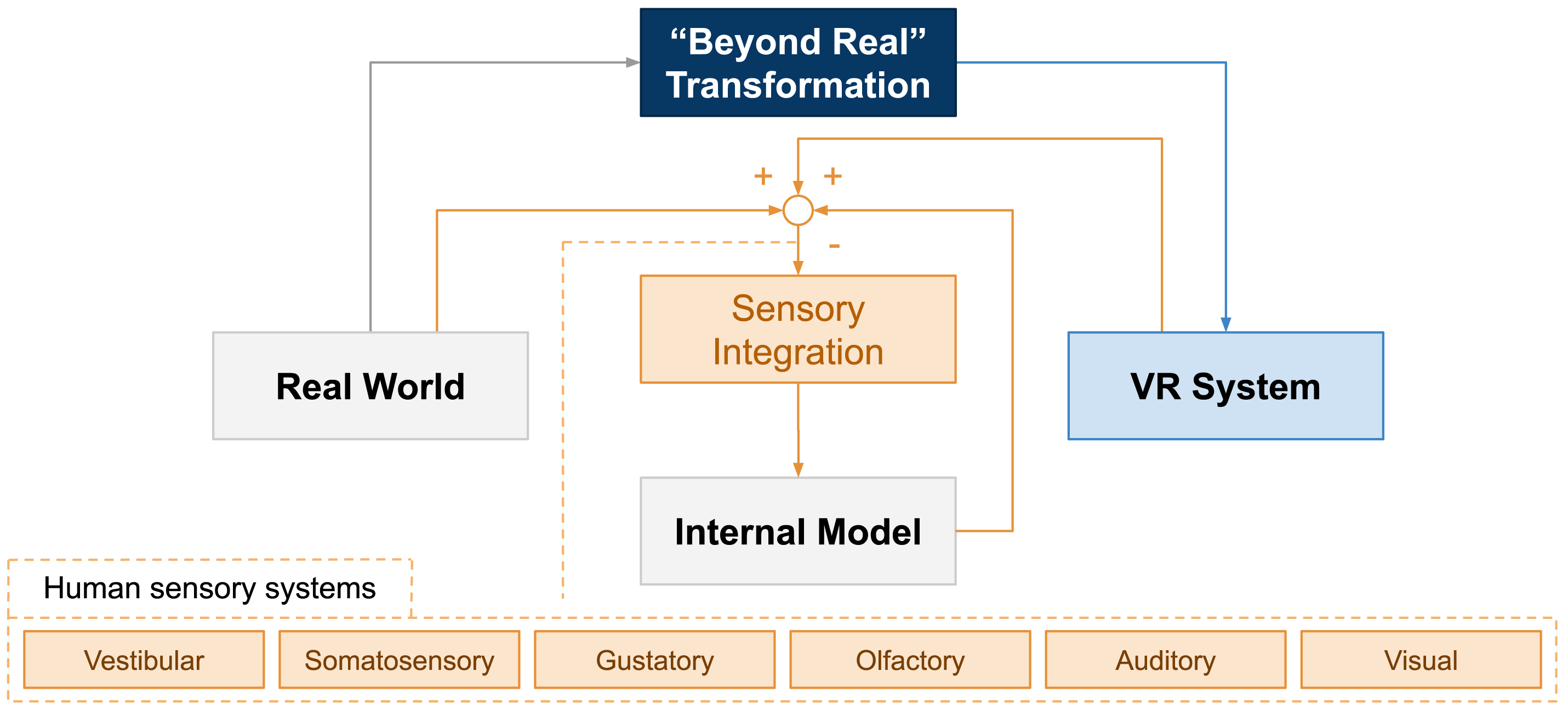}
        \caption{Users receive sensory information from both the real world and the VR system which are then integrated. Understanding sensory integration \new{and how the user's internal model is updated accordingly} is integral for exploring open research questions around beyond-real VR interactions.}
        \label{integration}
    \end{figure}
    

    Our survey identified that over the years, a number of reality-based and beyond-real interaction techniques have been developed and presented at research conferences. While a clear research agenda has been articulated for reality-based interactions (see \cite{jacob2008reality}), as a research community, we are missing a road map for the design and development of beyond-real interactions. Our survey also identified that while there are many new beyond-real interaction techniques developed, only a small percentage of them consider \new{sensorimotor} control issues beyond motion sickness. Our proposed framework, \textit{beyond being real}, allows for a uniform way to describe such interactions from this \new{sensorimotor} control perspective, but lacks any predictive ability. We believe a research agenda is needed to help bridge this gap towards better understanding and modeling of beyond-real interactions. Here we discuss some of the open research questions for beyond-real interactions from this perspective. 

\subsection{Control and Usability}
    In the real world, \new{once learned,} we perform skilled motor actions automatically and without conscious awareness of how they are executed \cite{rasmussen1983skills, norman2013design}. Initially, beyond-real remappings are unfamiliar, as they have no counterpart in our physical reality, and the forward model cannot predict the \new{outcome} of motor actions. These unexpected prediction errors shift the user's attention in the virtual environment \cite{arbib2009tool, talsma2015predictive} and lead to breakdowns that require conscious reflection during interaction, as described by the Heideggerian notion of presentness-at-hand \cite{heidegger1996being, winograd1986understanding}. Motor learning can be described as an experience-driven, systematic update to the internal model that enables users to predict the \new{outcome} of motor commands and develop new control policies \cite{flanagan2003prediction} (see Figure~\ref{brain}). For beyond-real interaction designs to be usable \cite{nielsen1994usability}, users need to learn to synthesize movement under the new dynamics \cite{haith2013model}. Meaningful feedback plays a key role in these adaptations, as has been shown for body transformations in VR \cite{day2019examining}. However, the best methods for supporting these adaptations are not known.
    
\subsubsection{Drawing from Prior Experiences}
    The time it takes for users to learn beyond-real remappings depends on their familiarity with the motor task \cite{haith2013model}. To address \new{this} lack of familiarity, designers have leveraged users' prior experiences by using themes from science fiction literature, or more broadly books, movies, and other narratives \cite{mostafa2014poster}. Another approach is to design interactions that indirectly utilize skills users have already developed in the real world. For example, eye gaze as a mechanism for selecting distant objects leverages a skill we have developed as a result of making eye-contact with others during conversation \cite{jacob2008reality}.
    
 \subsubsection{Learning Timescales}
    Repeated interactions are needed for users to learn new control policies over time. For example, in a study where the range of motion of the participants' arms and legs were swapped, it took around 10 minutes for participants to learn the remapping and to utilize the range of motion of their avatar's body parts \cite{won2015homuncular2}. Motor learning is driven by different processes at multiple timescales and it often involves quick approximations, followed by slow adjustments that enable fine tuning \cite{heuer2013towards}. Beyond-real interactions change how we perceive the affordances of objects and properties of the world around us. For example, body scaling influences users' perception of size and distance \cite{van2011being, kokkinara2015effects, banakou_illusory_2013}. This \emph{movement context} is captured by the internal model, and compared to the state updates, it changes at a much slower time-scale \cite{wolpert2000computational}. Moreover, ``motor learning undergoes a period of consolidation, during which time the motor memory is fragile to being disrupted''; therefore, dynamic remappings that may interfere with one another should not be presented in quick succession. 

 \subsubsection{Exploration}
    The uncertainty associated with outcomes of motor action directly relates to exploratory behavior \cite{ueyama2014mini}. When users suspect that their knowledge of the environment could be improved, they make exploratory choices to increase their learning rate \cite{doya2008modulators}. While the specifics of these learning processes in the brain is not fully understood \cite{lee2014neural}, prior research has proposed different strategies for encouraging exploration. For example, providing lower quality visual feedback may increase uncertainty, promote exploratory risk-taking, and lead to more accurate internal models under the new dynamics \cite{heuer2013towards}. This approach is at odds with the sentiment of effective feedback in interaction design, and more research is needed to unpack this interplay. 
    
    \vspace{0.15cm} 
    \noindent Due to these gaps in our understanding, more predictive models are needed to determine the boundaries of usable beyond-real VR interactions and how far we can push those boundaries \cite{won2015homuncular2}. More specifically, new methods are needed for evaluating the usability of a beyond-real interaction design and predicting how much time is necessary for users to learn the remapping. 

    
\subsection{Long-Term Use and Aftereffects} 
    It has been shown that our cognitive system can adjust to repeated exposure to conflicting stimuli \cite{biocca1998virtual, cruse1990utilization}. This adaptation, which is driven primarily by the forward model \cite{bhushan1999evidence}, has been studied both in the context of illusions and interactions beyond reality. The VR user experience begins when users choose to engage with the virtual content and put on their headset and continues as users exit VR \cite{Knibbe2018}. Therefore, adaptation aftereffects that may carry over into the real world need to be carefully considered.  Aftereffects are often task-dependent \cite{arbib2009tool} and may also depend on the speed of the movement \cite{cruse1990utilization}. These adaptations also affect other aspects of how we perceive space \cite{kokkinara2015effects}. However, not all adaptations lead to aftereffects, and this may depend on the modality of the sensory feedback. For example, in a between-subjects study where users experienced either visual or proprioceptive distortions (created by vibration), the effects of proprioceptive adaptations disappeared afterwards \cite{bernier2007sensorimotor}. Many research questions remain unanswered in the context of long-term use of beyond-real interactions. Can we train users, through extended practice, to perform effectively under novel remappings? How will the dynamics of the interaction change after the novelty of the beyond-real experience diminishes? Will users maintain their ability to perform under similar remappings the next time they return to the virtual experience?
    
\subsection{Individual Differences} \label{sec:individual}
    Individual differences play a significant role in how users perceive and \new{act} in virtual environments \cite{gonzalez2019individual}. \new{These differences also influence sensory integration and the thresholds at which users become aware of novel remappings. As a result, the categorization of interactions as either illusory or beyond-real is user-dependent. For example, a user that notices a reach redirection illusion {\cite{azmandian2016haptic}} will perceive this interaction as a beyond-real space transformation.} In the context of beyond-real VR interactions, various factors may contribute to these individual differences, including users' age \cite{heuer2010effects}, physique \cite{bowman1997evaluation}, prior experiences \cite{stepanova2019space, white1998overview}, familiarity with science fiction \cite{mostafa2014poster}, or gaming frequency \cite{witmer1998measuring}.
    
\subsubsection{Physiological Responses}
    Beyond-real interactions are susceptible to negative physiological responses, as users often receive incongruent sensory feedback from the real world and the virtual system. For example, some users report symptoms of motion sickness when there is a mismatch in sensory feedback from the vestibular and visual systems. Individuals may have different physiological responses to a virtual experience. With regards to flying, researchers have found that ``a lot of people find it an endless source of fun, but other people report tired arms and motion sickness'' \cite{bricken1991virtual}. The user's physiological response is an important consideration from both safety and usability standpoints \cite{bailenson2018experience, fox2009virtual}.
    
\subsubsection{Emotional Responses} 
    VR has been recognized as a powerful, immersive media that can evoke strong emotions in users \cite{bailenson2018experience}. In particular, experiences beyond reality may lead to profound emotional responses, both positive \cite{gaggioli2016human, stepanova2019space}, often involving the feeling of awe \cite{chirico2016potential, keltner2003approaching}, and negative, such as the feelings of fear \cite{bourdin2017virtual}, distress \cite{steptoe2013human, seinfeld_offenders_2018}, and regret \cite{friedman2014method}. It has been shown that incongruent sensory stimuli result in negative emotions \cite{schreuder2016emotional}. Conversely, emotional responses influence multisensory processing in ways that can be reflected in action \cite{latta1994conceptual}. For example, expectations of high reward release dopamine in the brain, such that it may no longer operate as an optimal controller \cite{doya2008modulators}.
    
    \vspace{0.15cm}
    \noindent At a high-level, individual differences influence sensorimotor control and consequently, how users respond to beyond-real VR interactions. However, many open research questions remain regarding how individual differences should be accounted for in the design of such interactions. How can we capture individual differences from real-world input data? How might we better model user behavior based on individual user's interactions with the system over time? Can we offer adaptive, personalized experiences that account for individual differences? How might we then evaluate beyond-real interaction designs at a large scale?
    
\subsection{Presence and Plausibility}
    Presence, or \emph{place illusion}, has been defined as the psychological experience of \emph{being there}: ``the extent to which an individual experiences the virtual setting as the one in which they are consciously present'' \cite{slater1997framework}. Designers, often guided by implicit or explicit theories, seek presence in the hopes of improving other attributes of the virtual experience, including learning or task performance \cite{biocca1997cyborg}. However, researchers have found inconsistent results when studying the correlation between the sense of presence and such attributes \cite{slater1996immersion, slater1997framework}, which perhaps is expected, as these are influenced by many factors, including users' abilities and prior experiences. Due to individual differences, in practice, it may be challenging to evaluate the effects of beyond-real interactions on users' subjective sense of presence. Witmer and Singer \cite{witmer1998measuring} have proposed a collection of factors hypothesized to contribute to presence, including control, sensory, distraction, and realism factors. It should be noted that the realism factor does not require content that replicates reality, but relates to continuities, connectedness, and coherence of the virtual experience. 
    
    \emph{Plausibility illusion}, the illusion that what is occurring is actually happening \cite{slater2009place}, is another psychological dimension that has been attributed to realistic responses to virtual environments. Plausibility illusion also does not require physical realism and is related to causal relationships between the user's actions and the resulting sensations. While in this paper we do not discuss presence and plausibility directly, human sensorimotor control naturally lends itself to discussions around these contributing factors.
    
\subsection{Accessibility} 
    
    Beyond-real interactions must also be considered from the perspective of accessibility. Mott et al. \cite{mott2019accessible} discuss the potential in VR for increased ``interaction accessibility'' and equity for all people, including people with disabilities, given that in VR people can have abilities no person can experience in the real world, what they call ``superpowers'' (e.g., flying). \new{Sadeghian and Hassenzahl extend this concept of superpowers into a VR interaction design methodology {\cite{sadeghian_limitations_2021}}.} While assistive technology in the physical world has many limitations in terms of its ability to adapt, VR and specifically beyond-real interactions might support more adaptive and ability-based interactions \cite{wobbrock2011ability}. However, researchers have also warned about the potential to amplify differences in ability \cite{mott2019accessible} and that the inherent body-centric perspective of VR poses substantial issues for people with physical disabilities. Gerlig and Spiel \cite{gerling2021critical} highlight the importance of considering minority bodies while designing VR interactions and more importantly including people with disabilities in the design of new interaction paradigms. Additionally, while some VR accessibility research, including the papers we surveyed, focus on manipulating visual feedback (e.g., \cite{mirzaei_head_2021}), there are opportunities for exploration of beyond-real interactions through other sensory feedback, which would increase accessibility for blind and visually impaired VR users, who primarily access VR through audio \cite{seki2010training, connors2014action} or haptics \cite{zhao2018enabling, siu2020virtual}. 
    

\subsection{Ethical Implications}
    Slater et al. present a detailed discussion of the ethics of realism in virtual reality, and many of the discussion points are highly relevant to beyond-real experiences \cite{slater2020ethics}. Throughout the paper, we have also alluded to some ethical implications of beyond-real VR interactions, such as motion sickness; however, it is necessary to explicitly acknowledge the importance of ethical considerations from a sensorimotor control perspective. Beyond-real remappings may result in motor behavior changes during the interaction. For example, when embodying an avatar with a flexible tail-like appendage, users changed the way they moved their hips \cite{won2015homuncular}. When walking around a virtual environment as a giant, users took bigger steps in the real world \cite{abtahi2019m}. ``Presence in VR leads to absence in the physical world'' \cite{bailenson2018experience}; therefore, not accounting for behavior changes, especially with regards to users' movements in the real world, could have serious consequences. Moreover, long-term use of beyond-real VR interactions may have aftereffects that alter motor behavior in the real world \cite{bernier2007sensorimotor}. For example, in a study where users' virtual eye position was offset from the position of their real eyes, it was shown that, after removing the VR headset, participants' hand-eye coordination was altered, as evidenced by their inability to accurately point to a target \cite{biocca1998virtual}. Such sensorimotor adaptations pose safety concerns that need to be carefully considered.

\section{Limitations}
\label{sec:limitations}
    In the following section, we will acknowledge some of the limitations of our work and highlight opportunities for future work related to the study of beyond-real virtual reality interactions.
    
    \subsection{Completeness}
    The human sensorimotor system is incredibly complex and many details are not captured by the simplified model presented here. For example, multisensory integration happens at multiple stages, utilizing both bottom-up and top-down processes \cite{talsma2015predictive}, and the sensory signals involve considerable delays that are largely missing in our model \cite{wolpert2000computational}. 
    
    Moreover, our goal for conducting a survey was to apply the framework to selected work exploring beyond-real VR interactions. One limitation is our choice of query; a broader query may be needed for an exhaustive categorization. For example, some VR interaction papers, such as \cite{lilija2020put}, did not appear in our search because neither ``VR'' nor ``virtual'' was mentioned in the title or the abstract of the paper. Future work may also consider analyzing other venues, such as the journal of Virtual Reality, and including VR interactions that were not in the scope of our work, such as system control and symbolic input \cite{bowman20043d}.

\vspace{-0.15cm}
\subsection{Embodiment}
    Previous research has studied the sense of embodiment, which has been defined as having three components: the sense of self-location, the sense of agency, and the sense of body ownership \cite{kilteni2012sense}. Our work lacks a coverage of this extensive body of research and how beyond-real interactions may influence users' sense of embodiment in VR. For example, it has been shown that in arm-extension techniques the sense of body ownership declines as the length of the virtual arm increases \cite{kilteni_extending_2012}. The effects of beyond-real transformations on body ownership have mainly been studied in isolation, and our understanding of how different transformations might interact with each other in more complex scenarios is limited. Consider the beyond-real interaction technique Ninja Hands \cite{schjerlund_ninja_2021}, where the movement of the user's hand is mapped to multiple virtual hands in space. In Ninja Hands, these virtual hands are visually disconnected from the user's body (see Figure~\ref{collage}b). While prior research has shown that virtual limbs that are visually connected to the user's body increase the user's sense of body ownership (as measured through their physiological responses) \cite{tieri2015body}, it remains unclear if many connected limbs, as in Ninja Hands, would have a similar effect. Users' sense of body ownership, which is subject to individual differences \cite{marotta2016individual}, may have implications for learning and adaptation of beyond-real interactions. Further research is needed to unravel the effects of beyond-real transformations on embodiment, including body ownership, agency, and self-location. 
    
\subsection{Focus on Action}
    While we focus on VR experiences that require users to act on the world, applications that do not focus on action could also be beneficial. For example, beyond-real experiences can be utilized for demonstration in educational applications, as they provide a unique opportunity for learning abstract concepts \cite{bricken1991virtual}. They can facilitate transformative experiences that evoke strong emotions and elicit new insights \cite{gaggioli2016human}. Beyond-real experiences can ignite one's imagination and foster creativity \cite{lanier2017dawn} \new{and} encourage positive behavior changes that may even transfer to the real world \cite{bailenson2018experience}.
    
\subsection{Social Interactions}
    In our framework, we have taken an ego-centric approach, focusing on a single user's interactions; however, VR is well suited for socialization and collaboration. When describing how \emph{reality} is reflected in the word \emph{virtual reality}, Lanier highlights that ``VR functioned as the interstices or connection between people; a role that had been previously taken only by the physical world \ldots A reality results when a mind has faith that other minds share enough of the same world to establish communication and empathy'' (p. 240) \cite{lanier2017dawn}.
    
    Bailenson et al. describe techniques beyond reality that change the nature of social interaction in collaborative virtual environments, including manipulation of self representation, sensory capabilities, and the temporal/spatial context \cite{bailenson2004transformed}. They argue that in VR, unlike face-to-face interaction, the user's rendered behavior can deviate from their actual behavior. The system can leverage this characteristic to, for example, improve communication by altering the user's rendered behavior such that it mimics the nonverbal behavior of others, referred to as the Chameleon Effect \cite{chartrand1999chameleon}. \new{While beyond-real social interactions have been explored} \cite{roth_technologies_2019, roth_beyond_2018}, many research questions remain. For example, how might users with dramatically different scales interact \cite{xia_spacetime_2018}? How does that influence their perception of interpersonal distance?  
    
 \subsection{Other Sensory Modalities}
    Our work mainly addresses users' visual and somatosensory systems. However, the current state of VR technology enables rendering of audio, and perhaps in the future commercial VR headsets may be able to engage our other sensory input channels, such as olfactory \cite{nakaizumi2006spotscents}. Virtual experiences beyond reality are not limited to vision and touch, and can span other sensory modalities. For example, in VR, we may gain the ability to smell the scent associated with others, in ways that we could know when someone familiar enters a room without seeing them, or whether that person has previously been in the same room.
    
\subsection{Mixed Reality Spectrum} 
    Finally, while we have specifically focused on virtual reality, beyond-real interactions may be integrated into other experiences on the mixed reality spectrum \cite{milgram1995augmented}. For example, the Go-Go arm-extension technique has been applied in augmented reality to enable interactions with distant objects in the real world \cite{feuchtner2017extending}. While similar transformations may be used to describe such interactions, further study of the implications and considerations is needed.

\section{Conclusion}
    In this paper, we first described a simplified model of \new{the} control signal flow during movement-based interactions \new{and situated VR interactions within this model}. We explained how intent is converted to motor commands in the central nervous system \new{resulting} in movements in the real world. \new{These} movements are tracked by the VR system and transformed into virtual renderings. Users receive sensory feedback from both the real world and the virtual system. \new{In most cases, the brain operates as an optimal controller} and with the use of the state estimator, respond\new{s} accordingly to perform \new{the} intended actions in VR. Using this simplified model, we partitioned the space of VR interactions into reality-based, illusory, and beyond-real \new{based on the magnitude of the resulting sensory conflict}. We then presented \textit{beyond being real}, a framework for describing beyond-real interactions as a set of transformations applied to real-world input. We conducted a survey of prior HCI literature (at CHI, \new{IEEE VR}, VRST, UIST, and \new{DIS} conferences) with a focus on selection, manipulation, locomotion, and navigation in VR. \new{We applied our framework to extract and categorize the beyond-real transformations in these works and highlighted a gap:} research that carefully considers the consequences of sensory \new{conflict} resulting from \new{beyond-real} transformations. Lastly, we \new{discussed} challenges and opportunities for future research towards the goal of better understanding \new{and evaluating} interactions beyond reality.

\begin{acks}
    We would like to thank Jeremy Bailenson and Mar Gonzalez-Franco for their help and guidance. We also thank Tamara Munzner, Evan Strasnick, Cole S. Simpson, Darrel R. Deo, the Stanford HCI group, and the anonymous CHI reviewers for their valuable feedback.
\end{acks}


\bibliographystyle{ACM-Reference-Format}
\bibliography{references}


\end{document}